\documentstyle[aps]{revtex}
\begin{document}
\title{ Size-shrinking of deuterons in very dilute superfluid nuclear matter.}
\author{U. Lombardo$^{1,2}$, P. Schuck$^{3}$}
\address{
\begin{center}
\begin{flushleft}
~~~~~~~~~~~~~~~$^1$Dipartimento di Fisica, 57, Corso Italia, I-95129 Catania, Italy\\
~~~~~~~~~~~~~~~$^2$INFN-LNS, 44, Via S.~Sofia, I-95123 Catania, Italy\\
~~~~~~~~~~~~~~~$^3$Institut de Physique Nucl\'eaire,Universit\'e Paris-Sud,
F-91406 Orsay Cedex, France.\\
\end{flushleft}
\end{center}
}
\maketitle
\begin{abstract}
It is shown within the strong-coupling 
BCS approach that, starting from the zero-density limit of superfluid nuclear
matter, with increasing density deuterons first shrink before they start
expanding.
\end{abstract}

\centerline{PACS Numbers : 21.65.+f,74.20.Fg}
\vskip 1.0truecm
The crossover from Bose-Einstein Condensation (BEC) of bound fermion pairs
in the low density limit to superconductivity or superfluidity for higher
densities is a very actively pursued field of research, since for instance 
such phenomena may play a role in high $T_c$ superconductors \cite{RAN}.
In a recent paper we have shown \cite{SCH}, using the strong-coupling BCS 
approach, that also in nuclear matter such a crossover may take place.
Indeed deuterons are bound proton-neutron (p-n) pairs which may turn into 
p-n Cooper pairs at higher densities \cite{SCH}. In fact the
situation in superfluid nuclear systems with $\Delta / \epsilon_F
\approx 1/10$, where $\Delta$ is an average gap value and $\epsilon_F$ the
Fermi energy, rather resembles a strong coupling superconductor than the one
of ordinary metals. 

In this note we will continue our previous work \cite{SCH} and study the size 
of the deuterons or p-n Cooper pairs as a function of density.
It is expected on arguments of general grounds that with increasing density,
starting from the zero-density limit, the deuterons first shrink before they
expand \cite{PIE}. Indeed once the deuterons come on average close enough so
that they start feeling the Pauli principle with their immediate neighbors
they can avoid the increasing repulsion in reducing their spacial extension. 
Of course this cannot go on for ever and soon the deuterons will
start overlapping, loosing their binding, and increase in size. It is very 
interesting that such a general feature is already contained in the BCS
approach to superfluidity \cite{PIE}. Therefore we will present calculations
of the coherence length of SD p-n pairing in symmetric nuclear
matter based on the strong-coupling BCS approach. 
The interaction adopted in the gap
equation is the Paris force projected into the SD channel, which reproduces
quite well the experimental phase-shifts of p-n scattering as
well as the deuteron binding energy. The single-particle energy 
$\epsilon (p) = p^2/2m + U(p)$ is calculated from the same Paris potential 
in the BHF approximation. The details of the procedure for
solving the BCS equations are reported elsewhere \cite{SCH}.
The coherence length is defined by
\begin{equation}
\xi^2 \,=\, \frac{\int d{\vec r} r^2 |\psi(\vec r)|^2}{\int d{\vec r} 
|\psi(\vec r)|^2} \,=\, \frac{\int d{\vec p} |\nabla\tilde\psi(\vec p)|^2}
{\int d{\vec p}|\tilde\psi(\vec p)|^2}  
\end{equation}
where $\psi(\vec r)=<c^{\dagger}(\vec r)c^{\dagger}(0)>$ is the pairing 
function and $\tilde\psi(\vec p)$ is the Fourier transform. Integration in 
momentum space is more suitable since the gap equation is solved in momentum 
space giving $\Delta(\vec p)$ and then
\begin{equation}
\tilde\psi(\vec p) \,=\, \Delta(\vec p)/ 
\sqrt((\epsilon (p) - \mu )^2 + \Delta(\vec p)^2).  
\end{equation}
In Fig.~1 the coherence length is plotted as a function of the density: the
crosses correspond to using free single particle spectrum,and the stars to
using BHF mean field. The values of $\xi$ tend to coincide at low density where
the mean field is negligible. At vanishing density $\xi$ approaches the 
deuteron radius and then it decreases according to the general trend. The
minimum at $\rho \simeq 0.036 fm^{-3}$ amounts a deuteron radius of less
than $2~fm$ and to an  interparticle distance of
about $3~fm$. Above the minimum the coherence length rapidly rises in
qualitative agreement with the weak-coupling prediction,i.e. $\xi \sim 
k_F/\Delta$.   

The above mentioned size-shrinking can be traced back to the Pauli blocking 
effect. A simple explanation starts from the Schr\"odinger-like equation
\begin{equation}
(p^2/m -2\mu) \psi (p) \,=\, (1 -2n(p))\sum V(p,p') \psi (p')
\end{equation} 
which is equivalent to the BCS equations \cite{NOZ,SCH}. At zero-density this
is the Schr\"odinger equation for the p-n system: the eigenfunction
$\tilde\psi_{0}$ corresponding to the energy eigenvalue $2\mu_0 = -2.2 MeV$ 
is the 
deuteron groundstate. At very low density one may treat the potential 
$U(p,p')=-2n(p)V(p,p')$ as a small 
perturbation and then estimate the energy correction for $\mu$ positive
close to zero by
\begin{eqnarray*}
\delta E \,=\, \frac{ \sum\sum \tilde\psi_{0}^* (p) U(p,p') \tilde\psi_0 (p')}
{\sum |\psi^{0}(p)|^2} \,=\,\mu_0 \frac{16}{\pi} \frac{p_F}{p_0} 
(1 - \frac{p_0}{p_F} 
arctg(\frac{p_F}{p_0}) 
\end{eqnarray*}
where $p_0 = \sqrt (-2m\mu_0)$. This energy correction makes the system more
bound. One may also estimate from Eq.~(4) the correction to the coherence 
length, which results in
\begin{equation}
 \xi \,\approx \frac{\hbar}{\sqrt{2} p_0} ( 1 - \frac{8}{3\pi} \frac{p_F^3}{p_0^3})
\end{equation}
It is worth noticing that this result is in agreement with the one of Ref.~[3],
 where
also a linear dependence of $\xi$ on the density was found with zero-range force.
In the derivation of Eq.~(5) no assumption has been made on the force except
that it gives a bound state at zero-density. 

At this point it may be appropriate to discuss under which circumstances one
may find a Bose-condensed gas of deuterons or a transition from p-n Cooper 
pairs to a Bose-Einstein condensation of deuterons. 
In the first place one may think of the far tail of density of heavier 
$N \simeq Z$ nuclei like they are or will be produced in the new exotic nuclear
beam facilities. In a region of densities $\frac{\rho}{\rho_0} \simeq 
\frac{1}{20}$ the radial distance from the centre of heavier nuclei is such
that deuterons with $r.m.s.$ of $ \simeq 2 fm$ (see our figure) can be easily  
accommodated in a Bose condensed state. This picture demands the validity of 
the local density approximation and to neglect quantum fluctuations. Both
approximations are, of course, questionable for finite nuclei but we know by
experience that always something remains in a more correct treatment of those
very simplifying assumptions. In this respect it could be very interesting to
trigger on very peripheral nuclear reactions and to measure the yields of
deuterons (or correlated p-n T=0 pairs) with respect to uncorrelated nucleons.
Also in expanding nuclear matter as it is produced in the late stage of central
collisions with energies of $E/A \sim \epsilon_F$ condensation phenomena in 
very low density nuclear matter could play a role. In this respect it should 
be noted that for densities where the chemical potential of the p-n pairs is
negative, i.e. where there is Bose-Einstein condensation of deuterons the
influence of the Pauli principle due to additional neutrons (asymmetric case)
should be minimal. This has been confirmed by recent numerical calculations
\cite{SEDR}.

In conclusion we have shown that in a very diluted superfluid gas of deuterons
the deuterons as a function of density first shrink by $\sim 35\%$ before 
they start expanding
again. This relatively large effect is due to the Pauli principle which is fully respected in the
BCS approach. The investigation has been performed at $T=0$. The extension to
finite temperature and to asymmetric matter is on the way.

\bigskip
\bigskip

\begin{center}
FIGURE CAPTION
\end{center}
\begin{center}
\parbox{12cm}{\small
Fig.~1 \ Coherence length vs density in BCS approximation. }
\end{center}\normalsize
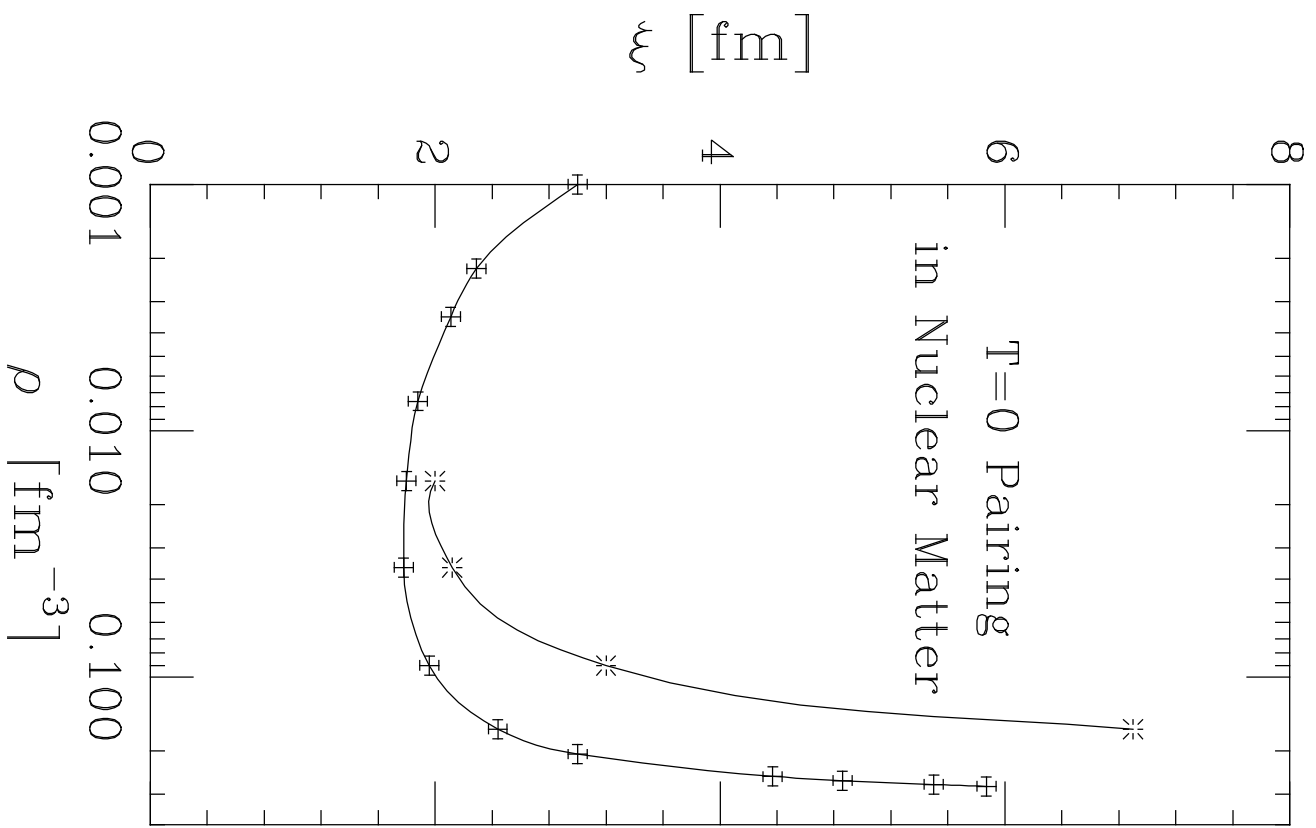
\end{document}